\documentstyle[times, emulateapj]{article}
\newcommand{\PSbox}[3]{\mbox{\rule{0in}{#3}\includegraphics{#1}\hspace{#2}}}

\begin{document}

\title{X 1908+075: An X-ray Binary with a 4.4 day Period}
\author{Linqing  Wen, Ronald A.  Remillard, and  Hale V. Bradt}
\affil{Center for Space Research, MIT, Cambridge, MA 02139 USA\\Email:  lqw@space.mit.edu, rr@space.mit.edu, hale@space.mit.edu}
\authoremail{ Email: lqw@space.mit.edu, rr@space.mit.edu, hale@space.mit.edu}

\begin{abstract}
X 1908+075 is an optically unidentified and highly absorbed X-ray source that
appears in early surveys such as Uhuru, OSO-7, Ariel V, HEAO-1, and
the EXOSAT Galactic Plane Survey. These surveys  measured a source intensity in
the range of  2--12 mCrab at 2--10 keV, and the position was localized to
$\sim 0.5 ^{\circ}$. We use the Rossi X-ray Timing Explorer ({\it
RXTE}) All Sky Monitor (ASM) to confirm our expectation that a particular Einstein IPC detection (1E
1908.4+0730) provides the correct position
for X 1908+075. The analysis of the coded mask shadows from the ASM for the
position of 1E 1908.4+0730 yields a persistent intensity $\sim 8$
mCrab (1.5--12 keV) over a 3 year interval beginning in 1996
February. Furthermore, we detect a period of 4.400 $\pm$ 0.001 days
with a false alarm probability $< 10^{-7}$. The folded light curve is
roughly sinusoidal, with an amplitude that is 22\% of the
mean flux. The X-ray period may be attributed to the scattering and
absorption of X-rays through a stellar wind combined with the orbital
motion in a binary system.  We suggest that X 1908+075 is an
X-ray binary with a high mass companion star.

\end{abstract}

\section{Introduction}
                   
X-ray surveys have
consistently indicated the existence of a moderately bright X-ray
source near the galactic longitude 41.97$^{\circ}$ and
latitude $-0.80^{\circ}$.  This source, which we refer to as X 1908+075,  was cataloged as 4U 1909+07 by Uhuru, the first earth-orbiting
mission dedicated to celestial X-ray astronomy
(\markcite{forman78}Forman et al. 1978). Subsequent determinations were cataloged as 1M 1912+077 with MIT/OSO 7 (\markcite{markert79}Markert et al. 1979), 3A 1907+074 with Ariel 5 (\markcite{warwick81}Warwick et al. 1981; \markcite{bell84}Bell Burnell \& Chiappetti 1984), 1H 1907+074 with HEAO A-1/LASS (\markcite{wood84}Wood et al. 1984), GPS
1908+075 with EXOSAT/ME (\markcite{warwick88}Warwick et al. 1988).  There are large uncertainties in the position determinations of these survey instruments, as shown in Fig. \ref{pos}. 
The HEAO A-2 experiment on HEAO-1 scanned this region and detected X-ray emission of $\sim 13$  mCrab in the 1.5--20 keV band (HEASARC archival data).  The source was also detected with the modulation collimator experiment (HEAO A-3) on HEAO-1, with an 8$\sigma$ detection in each of the two collimators for the energy range of 3--13 keV (HEASARC archival data).  The HEAO A-3 instrument  yields a multiplicity of relatively precise error regions which appear as a grid of diamonds in Fig \ref{pos}.   

The
Einstein imaging proportional counter (IPC) imaged
a portion of this region and detected a  source, 1E 1908.4+0730, the brightest source in the center of the field, at a significance level of $\sim  7\sigma $
 (HEASARC archival data). The IPC flux is about 0.3 mCrab (0.5--3.5 keV), possibly highly attenuated by interstellar absorption (see table 1).  The position (J2000) is $ \alpha =19^h 10^m 46^s$ and $\delta =+07^\circ 36'
07''$ with a positional accuracy of $50''$.    This position overlaps one of the HEAO A-3 ``diamonds''.   There are a few other 5-$\sigma$ detections at the edge of the field of view ($1^\circ$ square) of the IPC, and these sources are located well beyond the celestial map provided in Fig. \ref{pos}. Furthermore,  none of these sources have positions that are consistent with the HEAO A-3 diamonds.  It is thus highly likely that the Einstein/IPC detection represents the
same source as the X-ray survey detections.  No optical
counterpart has been identified within the error circle of 1E 1908.4+0730.

\setcounter{figure}{0}
\refstepcounter{figure}
\PSbox{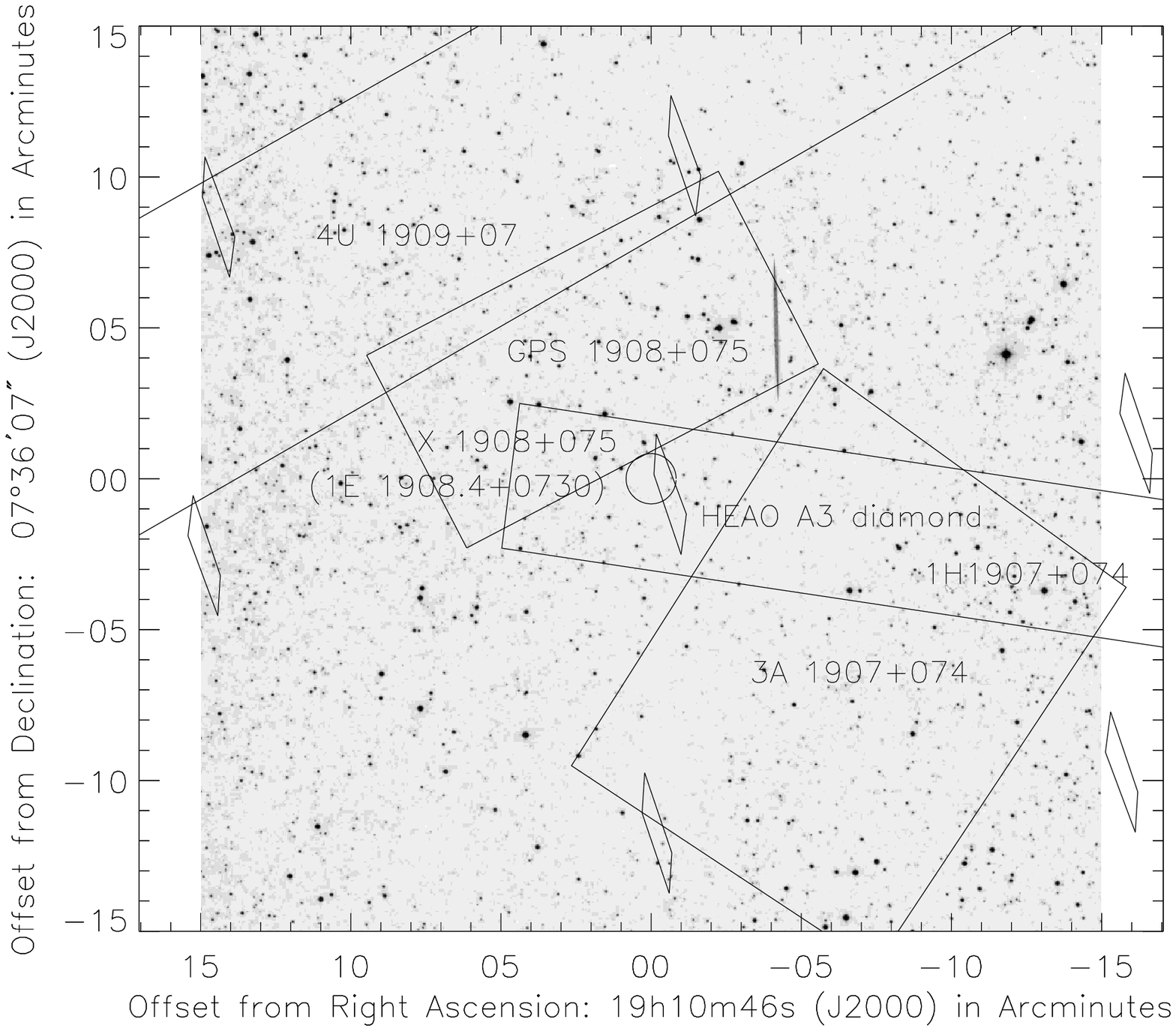 hoffset=-15 voffset=-5 hscale=45 vscale=45}{3.0in}{3.0in}{\\\\\small Fig.1  Positions of X 1908+075 and other X-ray sources detected previously, with reported error boxes superposed on an optical image taken from  the Digitized Sky Survey.  The degenerated positions determined with HEAO A-3 can be seen as a grid of diamonds.  The position of X 1908+075 used in the ASM analysis was taken to be that of Einstein/IPC source 1E 1908.4+0730, represented as a small circle.
\label{pos}}

\vspace{0.5cm}


The X-ray spectra and flux available from the hard X-ray survey detections are
 roughly consistent with each other as summarized in table 1. The fluxes are about 5 mCrab in the 2--6 keV band and 7--14 mCrab in the 2--20 keV band. The spectrum of EXOSAT/ME observation of
 4U 1909+07 in the $1$--$15$  keV band (at the UHURU position) indicates excessive absorption
 at photon energy below $6$ keV compared with that of the Crab Nebula (HEASARC archival data). The
 observations from the Ariel 5 spectrometer show that 3A 1907+074 has a hard spectrum
 which can be fitted with a power law of photon index $1.6 \pm 0.15$,
 hydrogen column density of $3^{+1.5} _{-1.2} \times 10^{22}$
 cm$^{-2}$, and an iron line emission with equivalent width $660 \pm
 220$ eV (\markcite{bell84}Bell Burnell \& Chiappetti 1984). None of these pointed observations were long enough to investigate
 the long-term variability of this source.

In this paper, we use the data from the All Sky Monitor (ASM) (\markcite{levine96}Levine et al. 1996) on board {\it RXTE} (\markcite{bradt93}Bradt, Rothschild, \& Swank 1993) to confirm the position and to study the nature of X 1908+075. We show that the coded mask patterns for the ASM,  processed for the
 Einstein/IPC position, yield a persistent $\sim 8$ mCrab X-ray
 source with a coherent $4.4$-day periodicity for over 3
 years. We then discuss the implications of
 this discovery.

\section{Observation}
The {\it RXTE}/ASM has been monitoring the sky
routinely since $1996$ February.  It consists of three Scanning Shadow Cameras, each
consisting of a coded mask and a position-sensitive proportional
counter. The ASM is well-suited for our purpose since each of its cameras  has a 1.5--12 keV response, an intrinsic angular resolution of a few arc minutes, and a large field of view of $6^\circ \times 90^\circ$(FWHM). In addition,   the ASM  provides observations with
relatively good sensitivity ($\le 5$--$10 $ mCrab in a day), frequent
data sampling ($\sim 5$--30 times a day) and long baseline (thus far 3
years).

For a 90-s exposure in each ASM camera, a set of source
intensities for the known sources in the FOV is derived from a $\chi ^2$ fit for the overlapping
shadow patterns of the X-ray sources and the X-ray background.  The intensity is computed in each of the three energy
bands ($1.5$--$3$, $3$--$5$ and $5$--$12$ keV) and the results are
normalized in units of the count rate for a source at the center of
the field of view of camera \# 1. In these units, the
$1.5$--$12$ keV flux for the Crab Nebula is $75.5$ ASM ct/s. The
estimated errors of the source intensities include the uncertainties
due to counting statistics and a systematic error of $1.9\%$
determined empirically from the light curves of the Crab Nebula. A
detailed description of the ASM and the light curves can be found in
Levine et al. (1996) and Levine (1998).

\section{Analysis and Results}

The X-ray light curve from the ASM observations of X 1908+075 (1996 March to 1999 January) is shown in Fig. \ref{ltc_all}.  The intensity points are 5.0-d average of the intensities derived from the individual 90-s dwells for the assumed position of the Einstein/IPC source.  It is apparent that X 1908+075 has generally
persisted at an intensity level of about 0.6 ct/s, or 8.0 mCrab in the 1.5-12 keV
band for about three  years.  The spatial  response function of the ASM implies that these photons arise within 12 arcmin of the Einstein/IPC source position, thus confirming its association with X 1908+075. The average X-ray intensity of X 1908+075 is well below the detection threshold (30 mCrab) for a single camera exposure  of the ASM. The effort to determine an independent source position using ASM data alone, e.g., with the superposition of cross-correlation maps,  would be limited by the appearance of systematic  noise feature with amplitudes that are a substantial fraction of the average source intensity.


The average values of the hardness ratios of X 1908+075 are HR1=$1.0 \pm 0.2$ , HR2=$3.6 \pm 0.3 $ respectively, where HR1 is defined as the ratio of the ASM count rates in the $3$--$5$ keV band to that of the
$1.5$--$3$ keV band, and HR2 is the ratio of the count rates in the
$5$--$12$ keV band relative to the $3$--$5$ keV band. 
 X 1908+075 has a much  harder X-ray spectrum than either the Crab Nebula (HR1=0.9, HR2=1.1) or Cyg X-1 in the low-hard state (HR1=1.0, HR2=1.5).

\refstepcounter{figure}
\PSbox{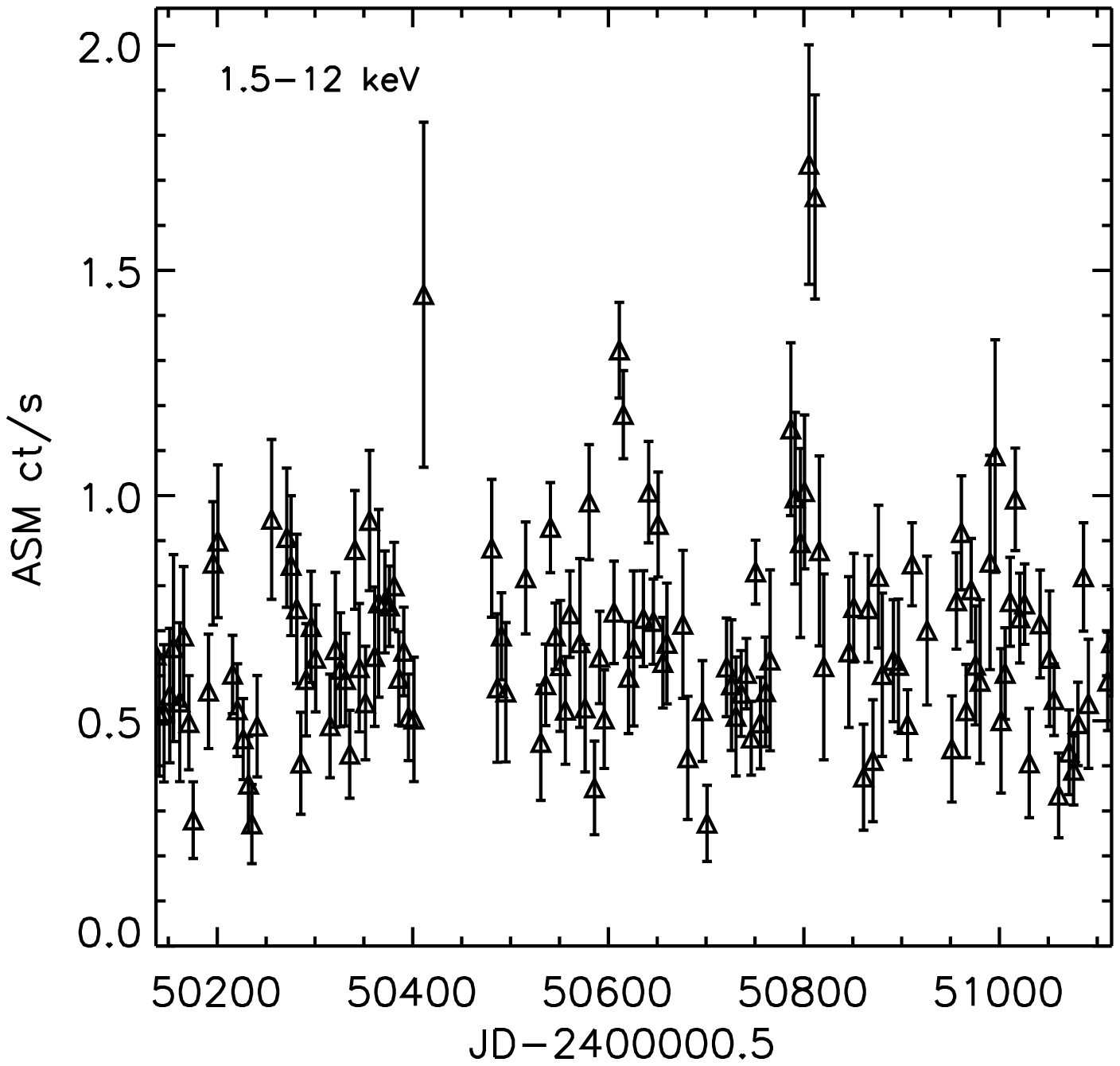 hoffset=-10 voffset=0 hscale=52
vscale=52}{3.2in}{3.2in}{\\\small Fig. 2 $\it RXTE$/ASM light curves of X 1908+075 for the
period $1996$ February to $1998$ November.  The data are binned in $5.0$-day bins.  Only $5\sigma$ detected data points are plotted. MJD $51000$ (MJD=JD-2400000.5) corresponds to
$1998$ July $6$. There is a persistent X-ray emission at $\sim 0.6$ ct/s (8 mCrab) for  3 years. \label{ltc_all}}
\vspace{0.5cm}

Periodicities of X 1908+075 were sought by means of the Lomb-Scargle
periodograms of both the light curves and the derived hardness ratios.
The Lomb-Scargle periodogram (\markcite{lomb76}Lomb 1976; \markcite{scargle82}Scargle 1982; \markcite{press92}Press et al. 1992) was used for the calculation instead of the Fast 
Fourier Transform (FFT) since the ASM data points are unevenly spaced
in time.  With this periodogram, a maximum in the power occurs at the
frequency which gives the least squares fit to a sinusoidal function. The expected average power for random noise is $1.0$. We
oversampled the power spectrum so that the frequencies are a factor of
4 more closely spaced than $1/T$, where $T$ is the total duration of
the data used.  The goal is to ensure the detection of a peak for a
signal that is of marginal statistical significance and to best
locate the peak.  The result is shown in Fig. \ref{power}.

There is a distinct narrow peak in the power spectra at a frequency
that is corresponding to $4.400$ d.  The width of the power peak is
0.009-d, consistent with that of a pure sinusoidal wave. The
uncertainty of this period is estimated to be $0.001$ d, assuming a
single signal with Gaussian noise. With 3 years of source coverage,
the uneven data spacing would not degrade the uncertainty to any
noticeable degree (\markcite{horne86}Horne \& Balinuas 1986).  If we assume the noise follows an
exponential distribution as expected for random noise, the false-alarm
probability of this detection is less than $ 10^{-7}$. In addition,
there is a low-frequency peak corresponding to about $180$-d
period. Except for these two peaks the overall power is flat with an
average of about $1.06$.

We have also conducted another independent periodicity search using
the epoch folding technique (\markcite{leahy83}Leahy et al. 1983; \markcite{davis90}Davies 1990).  The
data were folded modulo a trial period and then grouped by phases.  The resulting deviation from uniform random noise can
then be detected using a $\chi ^2 $ test.  For periods between 0.01--7
days, a distinct peak indicating the maximum deviation was found
around $4.400$ d with $< 10^{-6}$ false alarm probability level. This
is consistent with the results of the Lomb-Scargle periodogram.

\refstepcounter{figure}
\PSbox{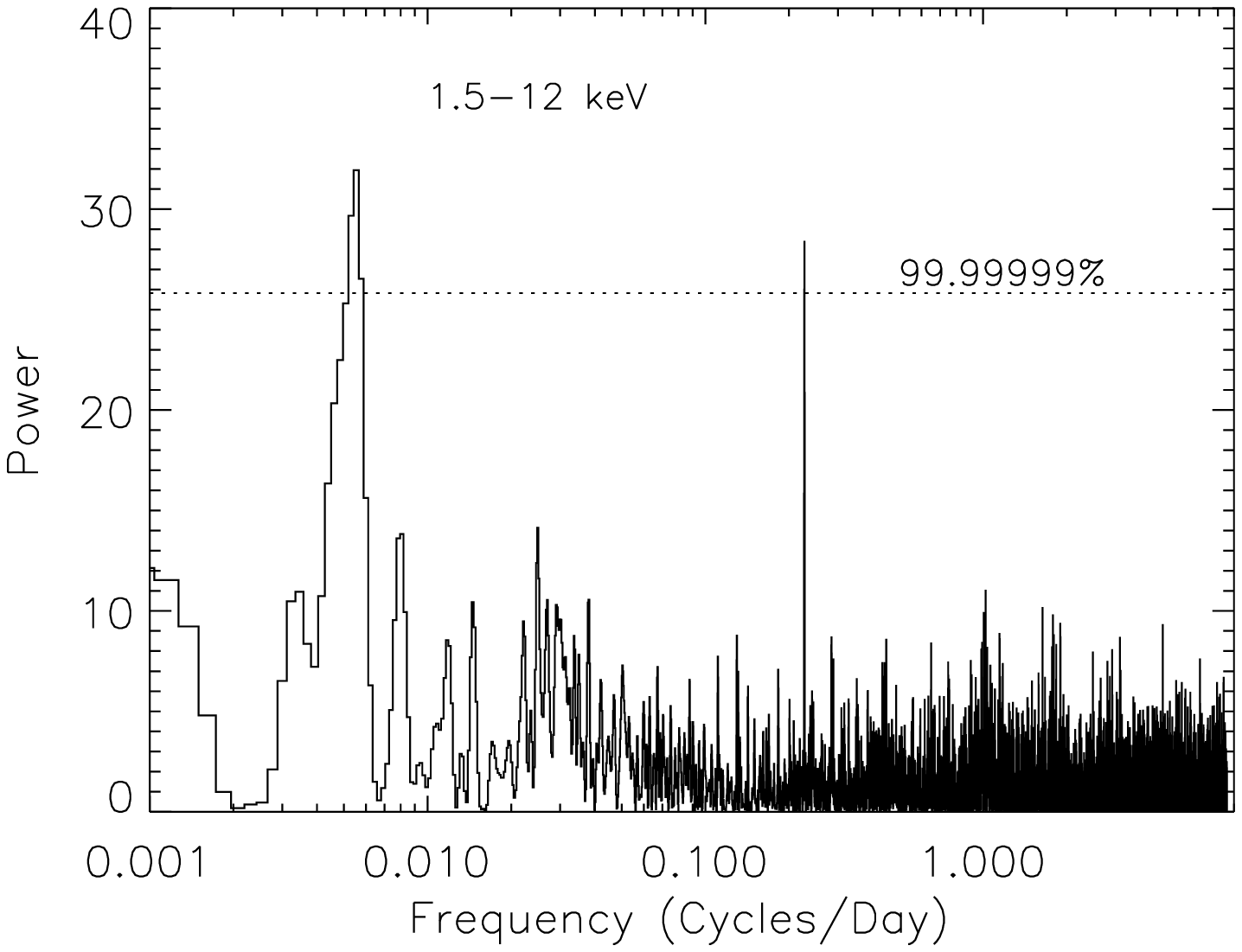  hoffset=-15 voffset=-30 hscale=55 vscale=60}{3.0in}{3.0in}{\\\small Fig. 3 The Lomb-Scargle periodogram of the light curves of X 1908+075. The dotted line indicates the $99.99999\%$ confidence level assuming an exponential distribution of the noise powers. There is a distinct peak at $4.4$-d period. The 180-d period is probably caused by scattered solar X-ray contamination. \label{power}}

\vspace{0.5cm}

The data were then folded modulo the period of the best-fit value of
$4.39986$ days from the Lomb-Scargle periodogram to study the
intensity variations with phase (Fig. \ref{fold}). 

\refstepcounter{figure}
\PSbox{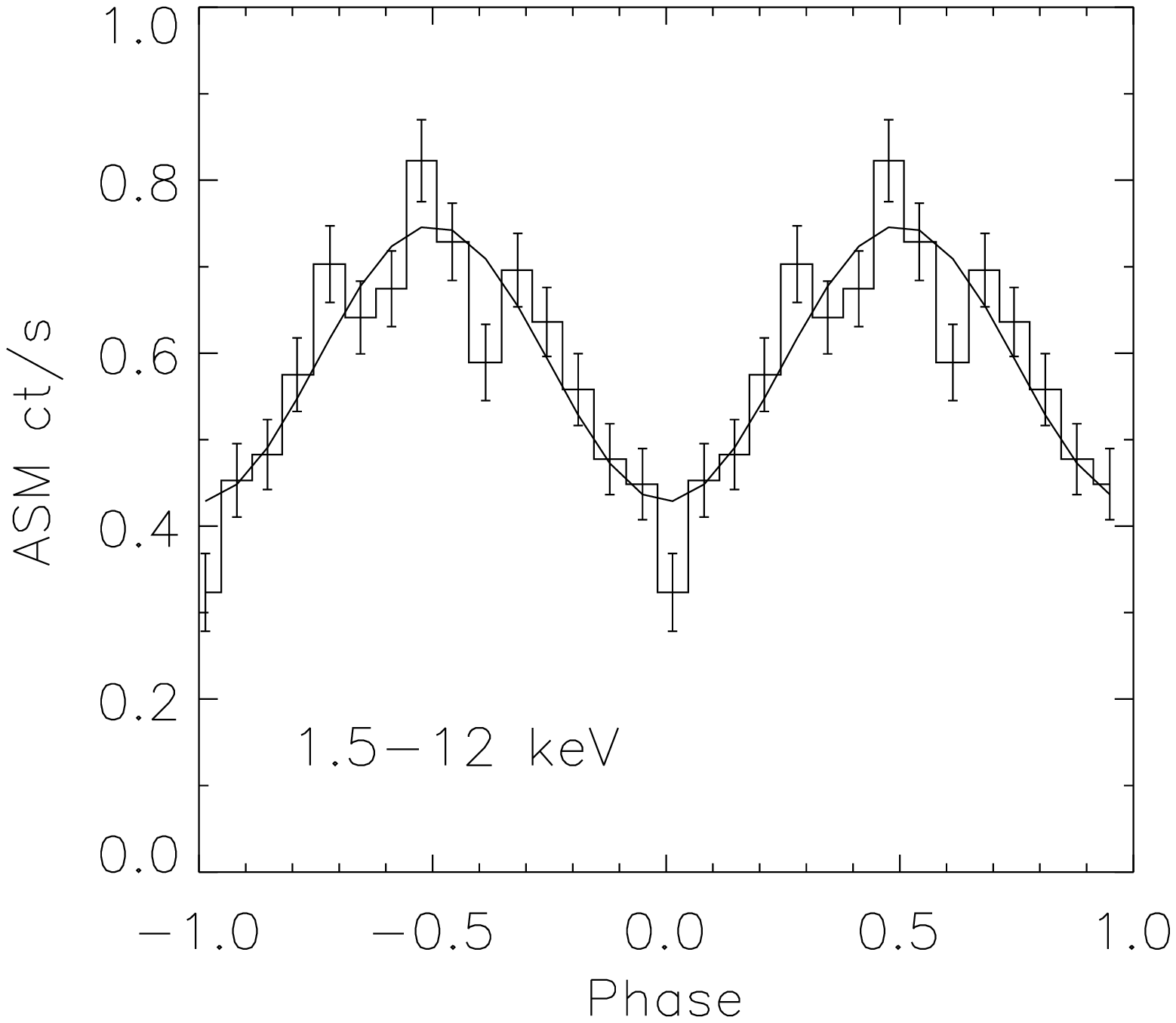 hoffset=-18 voffset=-15 hscale=50
vscale=50}{3.0in}{3.0in} {\\\\\small Fig. 4 Folded light curves of X 1908+075.  Observational data points are represented as histograms. Phase zero is defined as the phase of the minimum X-ray flux. The error bars represent one
standard deviation. The
solid lines are the best-fit sinusoidal waves. The most distinctive feature is the nearly sinusoidal modulation with large amplitude. \label{fold}}

\vspace{0.5cm}

 The most
distinctive feature in the folded light curves is the nearly  sinusoidal
modulation of large amplitude.  We fitted a sinusoidal wave plus a constant to the
folded light curves. The best fit yields a $26 \pm 3 \%$ amplitude of
the sinusoidal wave in the $1.5$--$12$ keV band, $51 \pm 14\%$ in the
$1.5$--$3$ keV band, $49 \pm 8\%$ in the $3$--$5$ keV band, and $22
\pm 3\%$ in the $5$--$12$ keV band. The best-fit epoch for the minimum
flux is:

\begin{center}
	$T_{0}=2450440.419 \pm 0.068 +n\times (4.400 \pm 0.001$ ) (JD)
\end{center}

The persistence and coherence of this periodic variation were tested
by subdividing the ASM light curves into four
consecutive $250$-d intervals and then searching for periodicities.  The same analyses described above were
applied to each of the $250$-d light curves. In the calculated power
density spectra, the signals are still visible as narrow peaks near
$P=4.40$ d in the power density spectra even though they are no longer
independently significant. The best-fit amplitudes and phases of the
sinusoidal waves are roughly consistent with each other (see Table
\ref{tab2} for a summary).  We thus conclude that the modulation is
persistent and coherent over the 3-year time interval.

\section{Discussion}

The X-ray emission detected by the ASM has been found to arise from the immediate region of the Einstein/IPC source, to exhibit a hard spectrum and intensity similar to the historical survey detections,  and to be persistent over $\sim 3$ years.  This together with the Einstein IPC and  HEAO A-3 positional correspondence strongly suggest that the position of 1E 1908.4+0730 is the
accurate location of the sources seen in the historical X-ray surveys.

It is natural to link the observed $4.4$-day period of X 1908+075 to
the orbital period of a binary. The periods of X-rays from known
sources are believed to be associated with one of the following
(\markcite{white95}White, Nagase, \& Parmar 1995): (1) rotational periods of neutron stars in X-ray
pulsars, ranging between milliseconds to hundreds of seconds; (2)
orbital periods of binary systems, ranging from tens of minutes to
tens of days, and possibly hundreds of days; and (3) ``superorbital''
orbital periods from $\sim 30$ days to hundreds of days, which in some cases are believed to be the precession period of the accretion
disk (e.g., \markcite{levine82}Levine \& Jernigan 1982).  It is clear that the coherence and the timescale of the $4.4$-day X-ray oscillation in X 1908+075 can be  best interpreted as the  orbital period of a binary system.

We can then picture X 1908+075 as a binary system consisting of a
compact object and a mass-donor companion star. An X-ray binary with a
low mass companion (LMXB) is unlikely the scenario for X 1908+075,
since the known orbital periods of most LMXBs (White et al. 1995) are on
the order of hours, much shorter than the detected 4.4 days. There are
only a few known LMXBs having orbital periods longer than a day, among
them only Cir X-1 ($P \sim 16.6$ d) and Her X-1 ($P \sim 1.7$ d, a
possible LMXB) exhibit orbital modulations in X-rays.  The modulation
in X-rays of Cir X-1 manifests itself as periodic flares and dips
likely due to a highly eccentric orbit, while Her X-1 is an eclipsing
system.  Such dramatic modulations are not evident in the orbital
light curves of X 1908+075. On the other hand, most HMXBs have orbital
periods on the orders of days.  The smooth and nearly sinusoidal
modulations in X 1908+075 have a lot in common with a supergiant
systems such as Cyg X-1.  We therefore suggest that X 1908+075 is a
high mass X-ray binary (HMXB).

We further suggest that the companion star is a supergiant rather than
a Be star for the following reasons.  (1) The $4.4$-d orbital period
fits in the range favored more by supergiant systems than by Be-star
systems, as the known orbital periods of the latter are much
longer than $4.4$ days ($\ge 16.7$-d) (White et al. 1995). (2) The known orbital light
curves of Be-star systems are all outburst-like rather than sinusoidal
(e.g., EXO 2030+375), probably due to the enhanced accretion caused by
the the compact object's periastron passage around the Be star.  (3) The orbital modulation of a supergiant system are known to be persistent while this is generally not true in 
Be-star systems (e.g., 4U 0115+63 in ASM light curves during 1996 March
to 1998 November).

The smooth and roughly sinusoidal orbital modulation in the X-rays of
X 1908+075 suggests that the system could be embedded in a strong
stellar wind from the companion star (e.g., a supergiant O or B star),
which absorbs and scatters the X-rays. The X-ray modulation is then
caused by changes in the optical depth along the line of sight to the
X-ray source as a function of orbital phase.  For X 1908+075,
absorption of X-rays probably dominates over scattering, as the
amplitude of the modulation in the $1.5$--$5$ keV band is larger than in the $5$--$12$ keV band.  To explain a modulation with an amplitude of  nearly $\sim 22\%$ in the 5--12 keV band, the difference in
the hydrogen column densities between phases 0 and 0.5 is expected
to be more than  $10^{23}$ cm$^{-2}$ for an absorption and scattering process.
The mechanism of the orbital modulation of X 1908+075 could be very
similar to that of Cyg X-1, wherein   hard state orbital modulation in the ASM light
curves can be well explained with
absorption and scattering in a stellar wind (\markcite{wen99}Wen et al. 1999).  Another
example of such a system could be the newly discovered X-ray pulsar
XTE~ J1855-026 (\markcite{corbet99}Corbet et al. 1999).  In this system, the orbital period
is proposed to be 6.1-d as detected by the ASM.  The modulation is
roughly sinusoidal with an amplitude about $22\%$ in the 1.5--12 keV band.  It has
also been proposed as a supergiant wind accretion system.
 
The hydrogen column density $n_{H}$ can be used to estimate the
optical extinction in the V band using the empirical relation between
the $n_{H}$ and the color excess $E_{B-V}$ (\markcite{savage79}Savage \& Mathis 1979).  It is
found that $A_{V}=15$ for $n_{H} =3\times 10^{22}$ cm$^{-2}$, as
indicated from the Ariel 5 observation of 3A 1907+074. The optical
objects within the Einstein/IPC error circle have magnitudes $V \ge
19$ in the Palomar sky survey.  To estimate the distance to X 1908+075,
we first investigated the possible range of the absolute V-band
magnitudes $M_V$ (from \markcite{lang92}Lang 1992) known for supergiant stars in
binaries with known orbital periods.  We found that $M_V \le -5.2$.
Together with $A_{V}=15$ and $V \ge 19$, we obtained $0.7$ kpc as the
lower-limit for the distance. The X-ray luminosity, estimated from the
ASM count rate (1.5--12 keV), is then $L_{1.5-100 keV} \ge 5 \times
10^{34}$ erg s$^{-1}$ if we assume the Ariel 5 spectrum. Given the
$A_V$ value, a more likely scenario is a distance $\sim$ 7 kpc
estimated from the formula provided by Allen (1973), where the
relation between the extinction of the star light near the galactic
plane and the distance is estimated based on the average properties of
interstellar absorbing clouds and the grains between them.  We
obtained similar distance by comparing directly the $n_H$ value of
X 1908+075 with that of another galactic plane source GRS 1915+105, which is
$3^\circ$ away from X 1908+075 and is believed to have a kinetic distance of 12.5 kpc
and $n_H = 5\times 10^{22}$ cm$^2$ (\markcite{chaty96}Chaty et al. 1996).  At the distance of 7 kpc,
the estimated X-ray luminosity of X 1908+075 is $L_{1.5-100 keV} \sim
5 \times 10^{36}$ erg s$^{-1}$.

Finally, the detected $180$-d period seems to be caused by scattered solar X-ray contamination.  We found from the ASM light curve that three major flares happened at the times when the sun was near the field of view of this source, while flares with smaller amplitudes are evident when the sun is  $180^\circ$ away from the source position. The latter may be caused by the back-scattering of the solar X-rays off  the earth's atmosphere and the instrument.

\section{Conclusions}  
 
A  persistent  hard X-ray emission ($\sim 8$ mCrab at 1.5--12 keV) of X 1908+075 has been detected with  the {\it RXTE}/ASM at the Einstein/IPC position of $ ra=19^h 10^m 46^s$, \ \ $dec=7^\circ 36'
07''$ (J$2000)$ for a 3-year  period. A  $4.4$-d period has been dicovered from the ASM light curves.  The modulation is nearly sinusoidal, persistent and coherent for  3 years. The amplitude of the best-fit sinusoidal wave is $26 \pm 3 \%$ in the $1.5$--$12$ keV band.  Our results support the simple interpretation that X 1908+075 is  the same source as 4U 1909+07,  3A 1907+074, 1H 1907+074, GPS 1908+075, 1E 1908.4+0730, as well as the source detected by HEAO A-3.  We argue that the 4.4-d period is the orbital period of a high mass binary system.  We also suggest that this system consists of a compact object and a supergiant star that produces a strong stellar wind. The X-ray modulation is then caused by changes in the optical
depth along the line of sight to the X-ray source as a function of orbital phase.

\acknowledgments 
We are very grateful to the entire $\it{RXTE}$ team at MIT for their support. Support for this work was provided in part by NASA Contract NAS5-30612.

\newpage


\clearpage
\newpage

\begin{deluxetable}{lccc}
\small
\tablecolumns{4}
\tablewidth{0pc}
\tablecaption{Historical Observations of X 1908+075 \label{tab1} }
\tablehead{Instrument & Observation Time & Energy Band & Count \nl &  & (keV) &(mCrab)
}
\startdata
RXTE/ASM &   1996-present & 1.5-12  & 7.7 $\pm$ 2.0 \nl
EXOSAT/ME &     1983 & 2-6  & $5.4 \pm 0.5$ \nl
Einstein/IPC & 1980 & 0.5-3.5 & 0.27$\pm$ 0.04 \nl
HEAO A2  &    1977 & 1.5-20 & 13.1 $\pm$ 0.1   \nl
ARIEL 5/SSI  & 1977  & 2-10 & 8 $\pm$ 4 \nl
UHURU &  1971--1973 & 2-6 & 4.9 $\pm$ 0.5 \nl
\tablecomments {Data were taken  from the HEASARC archive.  Each count rate was
normalized with respect  to that of
the Crab Nebula  detected by the same instrument in the specified energy
band. The errors were estimated based on the given values and scatter of
the measurements.}
\enddata
\end{deluxetable}

\clearpage
\newpage
\begin{deluxetable}{ccccc}
\tablecolumns{5}
\tablewidth{0pc}
\tablecaption{Sinusoidal Modulations of X 1908+075 \label{tab2}}
\tablehead{Time range\tablenotemark{a}& Fractional Amplitude &$\Delta 
\phi$\tablenotemark{b}& $\chi ^2 _{\nu}$\tablenotemark{c}\nl (d) & (\%)& & & 
}
\startdata
0--250   & 31.0 $\pm $ 7.4 & 0.046 $\pm$ 0.034 & 0.9 \\
250--500  & 25.9 $\pm $ 5.4 & -0.011 $\pm$ 0.032 & 0.8 \\
500--750  & 22.7 $\pm $ 10.0 & -0.065 $\pm $ 0.077 & 2.0\\
750--1000 & 30.0 $\pm $ 5.6 & 0.029$\pm$ 0.028 & 0.9\\ 
\enddata
\tablenotetext{a}{ The start time is 2450088.871 (JD)}
\tablenotetext{b}{ Difference between the best-fit phase $0$ and  the epoch given in the text. The error for interval 500--750 d was multiplied by a factor of 2  to take into account of the large value of $\chi ^2$}
\tablenotetext{c}{ Reduced Chi-square with $12$ degrees of freedom}
\end{deluxetable}

\end{document}